# Pressure-dependent semiconductor-metal transition and elastic, electronic, optical, and thermophysical properties of SnS binary chalcogenide

Ayesha Tasnim, Md. Mahamudujjaman, Md. Asif Afzal, R.S. Islam, S.H. Naqib*
Department of Physics, University of Rajshahi, Rajshahi 6205, Bangladesh
*Corresponding author; Email: salehnaqib@yahoo.com

**Abstract**

Density functional theory based study of the pressure dependent physical properties of binary SnS compound has been carried out. The computed elastic constants reveal that SnS is mechanically stable and brittle under ambient conditions. With increasing pressure, the compound becomes ductile. The Poisson's ratio also indicates brittle-ductile transition with increasing pressure. The hardness of SnS increases significantly with pressure. The compound possesses elastic anisotropy. The ground state electronic band structure is semiconducting with a small band gap which becomes metallic under pressure. The band becomes more and more dispersive with the increase in pressure while the electronic correlations decrease as pressure is raised. Both the Debye temperature and the phonon thermal conductivity of SnS increase sharply with pressure. The Melting temperature of the compound is low. Mixed bonding characteristics are found with ionic and covalent contributions. SnS is a good absorber of ultraviolet light. The reflectivity of the material increases with the increase in pressure. The reflectivity is nonselective over a wide spectral range. The low energy refractive index is high. All these optical characteristics are useful for prospective optoelectronic device applications. The optical anisotropy is low.

## 1. Introduction

Tin sulfide (SnS) is a chalcogenide binary compound that is a semiconductor at room temperature and ambient pressure. Two phases of tin sulfide are commonly found. They are $\alpha$-SnS-*Pnma* and $\beta$-SnS-*Cmcm*. The $\alpha$-SnS-*Pnma* phase is stable under ambient pressure, but it was found to undergo a phase transition to *β*-SnS-*Cmcm* under the application of pressure. After surveying many materials based on their possible photovoltaic applications, the sulfosalts were found to be the most promising candidates for future solar cell devices. Among these sulfosalts, due to their unique properties, tin mono sulfide was considered as one of the most suitable compounds for various applications, particularly for the fabrication of solar cell devices [1]. As one of the most popular metal sulfides, tin sulfide (SnS) was developed in the past few years as a safe and sustainable semiconducting material. Special attention has been attached to SnS because of having an optical band gap of the value of 1.38 eV [2], which is in between that of Si (1.12 eV) and GaAs (1.43 eV) and is close to the optimum value of 1.50 eV that is needed to allow the solar radiations to be efficiently absorbed. Additionally compared to the materials CdTe [3] and $CuInSe_2$ [4] that were being recently used, SnS has a higher absorption coefficient. Also, the elements that constitute SnS, namely tin (Sn) and sulfur (S), are present in abundance in nature and are non-toxic [5]. It is also estimated that the theoretical photovoltaic conversion efficiency of this compound is greater than 24% [6].

Application of pressure can tune the physical properties of solids. Pressure induced change in the electronic ground state is of particular interest [7,8]. For example the superconducting transition temperature shows strong pressure and hole content dependent variations in high-$T_c$ cuprates [9-11]. It is found in this study that SnS exhibits semiconductor-metal transition. SnS also exhibits superconductivity under high pressure in the metallic state under cooling.

As far as our knowledge is concerned some of the important mechanical, bonding, and optical properties of SnS remain yet to be studied carefully. In addition, the pressure-dependent studies of the structural, mechanical, electrical, optical, and thermophysical properties remain largely unexplored. We wish to bridge those gaps in our studies. These unexplored properties are very important for the understanding of the material for its possible applications. Besides, the SnS compound is a layered compound, this type of material possesses many exciting features [12-20] for which this particular compound demands attention and needs to be studied in detail.

In this paper, a detailed investigation of the mechanical, electronic, optical, and thermophysical properties of SnS has been conducted under different pressures. The mechanical properties include structural and elastic properties. The bonding characteristics are investigated. The optical properties are also studied carefully since SnS is used in the optoelectronic device industry. Possible role of pressure on superconductivity has been explored.

The rest of the paper has been arranged in the following manner: In section 2, the computational methodology that is used in this study is briefly described. In section 3, the crystal structure is described along with the computational results and their analyses. Finally, in section 4, the important findings of this study are discussed and summarized.

**2. Computational Method**

The density functional theory (DFT) is one of the most popular and practical approaches for the ab-initio modeling of structural and electronic properties of solids. Here, the solution of the Kohn-Sham equation [21] gives the ground state of the crystalline system. The quantum mechanical computational prescription implemented in the CAmbridge Serial Total Energy Package (CASTEP) [22] is used to carry out the DFT-based calculations. The electronic exchange-correlation in CASTEP that is used in this study is the Generalized Gradient Approximation (GGA). The GGA was chosen in the scheme of Perdew-Bruke-Ernzerhof (PBE) [23]. The Vanderbilt-type ultra-soft pseudopotential is used to model the Coulomb potential energy caused by the interaction between the valence electrons and ion cores. One of the advantages due to which ultra-soft pseudopotential is used is to save substantial computational time without affecting the accuracy appreciably [24]. To obtain the lowest energy crystal structure of SnS, geometry optimization is performed using the Broyden Fletcher Goldfarb Shanno (BFGS) minimization scheme [25]. The electronic orbitals used for S and Sn to derive the valence and conduction bands are S[$3s^2$, $3p^4$], and Sn [$5s^2$, $5p^2$]. The cutoff energy for the plane-wave expansion for SnS is set to be 350 eV. The Monkhorst Pack mesh [26] is used to carry out the sampling of the Brillouin zone (BZ) with a mesh size of 5×15×15 k-points in the momentum space. The geometry optimization of SnS is performed by using the total energy convergence tolerance of $5.0 \times 10^{-6}$ eV/atom, maximum lattice point displacement within $5.0 \times 10^{-4}$ Å, maximum ionic Hellmann – Feynman force within 0.01 eVÅ$^{-1}$ and maximum stress tolerance of 0.02 GPa with fixed basis set corrections [27]. Reliable estimates of structural, elastic, and electronic band structure properties with an optimum computational time are produced with these selected levels of tolerances. It was seen from various studies conducted in the past that the spin-orbit coupling (SOC) has small effects on the bulk physical properties such as the optimized lattice parameters, elastic constants, mechanical anisotropy, chemical bonding, thermos-physical behavior, and optical parameters a large number of metallic and semiconducting compounds [28-31]. As the main focus of this investigation is on the bulk physical properties of the given material, the SOC was not included in the calculations to follow.

The stress-strain method [32] is used to calculate the single-crystal elastic constants, $C_{ij}$ for the SnS system. SnS is an orthorhombic system. From the symmetry considerations an orthorhombic crystal has nine independent elastic constants ($C_{11}$, $C_{22}$, $C_{33}$, $C_{44}$, $C_{55}$, $C_{66}$, $C_{12}$, $C_{13}$, $C_{23}$). All other elastic properties such as the bulk modulus, *B*, shear modulus, *G*, and Young modulus, *Y* are also obtained from the calculated values of single-crystal elastic constants, $C_{ij}$ using the Vigot-Ruess-Hill (VRH) approach [33,34]. CASTEP also enabled the calculation of the imaginary part of the dielectric function, $\varepsilon_2(\omega)$, by using the formula given below:

$$\varepsilon_2(\omega) = \frac{2e^2\pi}{\Omega\varepsilon_0} \sum_{k,v,c} |\langle \psi_k^c | \hat{u}.\vec{r} | \psi_k^v \rangle|^2 \delta(E_k^c - E_k^v - E) \qquad (1)$$

where $\Omega$ represents the unit cell volume, $\omega$ represents the frequency of the incident photon, *e* denotes the electronic charge, and the unit vector defining the incident electric field polarization is represented by $\hat{u}$. $\psi_k^c$ and $\psi_k^v$ are the conduction and valence band electron wave functions at a given wave vector $k$, respectively. The calculated electronic band structure is used in this formula. Kramers-Kronig transformation equation is used to obtain the real part of the dielectric function, $\varepsilon_1(\omega)$ from the corresponding imaginary part, $\varepsilon_2(\omega)$ with the complex dielectric function expressed as:

$$\varepsilon(\omega) = \varepsilon_1(\omega) + i\varepsilon_2(\omega) \qquad (2)$$

The refractive index, the absorption coefficient, the energy loss-function, the reflectivity, and the optical conductivity are extracted from the obtained values of $\varepsilon_1(\omega)$ and $\varepsilon_2(\omega)$ [35]. The Debye temperature and other thermo-physical parameters are calculated from the elastic properties and elastic moduli of SnS.

To study and investigate the bonding characteristics of the solid for better understanding of the mechanical properties, the Mulliken bond population analysis has been used [36]. The projection of the plane-wave states onto a linear combination of atomic orbital (LCAO) basis sets [37,38] is used for SnS. The Mulliken density operator written on the atomic (or quasi-atomic) basis is used to implement the Mulliken bond population analysis,

$$P_{\mu\nu}^M(g) = \sum_{g'} \sum_{\nu'} P_{\mu\nu'}(g') S_{\nu'\nu}(g - g') = L^{-1} \sum_k e^{-ikg} (P_k S_k)_{\mu\nu'} \qquad (3)$$

## 3. Results and Discussion

### 3.1 Structural Properties

The crystal structure of SnS is orthorhombic. From a schematic view, the structure of SnS appears to be a slightly disordered NaCl-type structure. This is because the highly electronegative S atoms draw electron pairs from Sn and become $[Ne]3s_2 3p_6$ and $[Kr]4d_{10}5s_2 5p_0$. Additionally, it is also observed that the lattice experiences strong distortion due to the nonbonding 5s lone pair electrons of the Sn which causes the lattice to go from a rock-salt structure to a distorted orthorhombic layered structure. From **Figure 1** it can be seen how each Sn atom is coordinated by six S atoms with three short Sn-S bonds (~0.266 nm) within the layer, i.e., intralayer and three long Sn-S bonds (at distances ~0.338 nm) in the next layer, i.e., interlayer in these disordered-layered structures. In the case of both intralayer and inter-layer S atoms, the lines connecting the Sn with each other are found to be approximately mutually perpendicular. From this, it can be inferred that a weak van der Waal's force connects these layers along the *c*-axis. It is because of this that the layers of the SnS compound can be cleaved easily perpendicular to the *c*-axis.

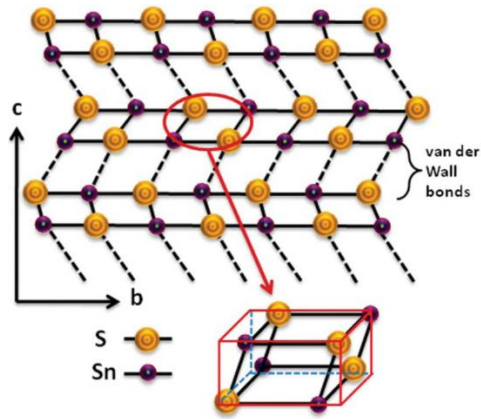

**Figure 1.** Schematic diagram of double-layered structured SnS.

Tin sulfide (SnS) crystallizes in the GeS (B16) type structure in the space group, *Pnma*. It is seen that Sn and S atoms occupied (0.1194, 0.25, 0.1198) and (0.8508, 0.25, 0.4793) positions respectively. The fractional coordinates of the atoms Sn and S in SnS are recorded in **Table 1.**

**Table 1.** Fractional coordinates of the atoms in the unit-cell of SnS.

| Compounds | Atoms | Fractional coordinates | | | Ref. |
|---|---|---|---|---|---|
| | | $x$ | $y$ | $z$ | |
| SnS | S | 0.8508 | 0.25 | 0.4793 | [39] |
| | Sn | 0.1194 | 0.25 | 0.1198 | |

As mentioned earlier, SnS crystallizes in the orthorhombic system in the space group *Pnma* (number-62) and that for SnS, the number of atoms per unit is 2. The three-dimensional schematic representation of the crystal structure of SnS is given in **Figure 2**.

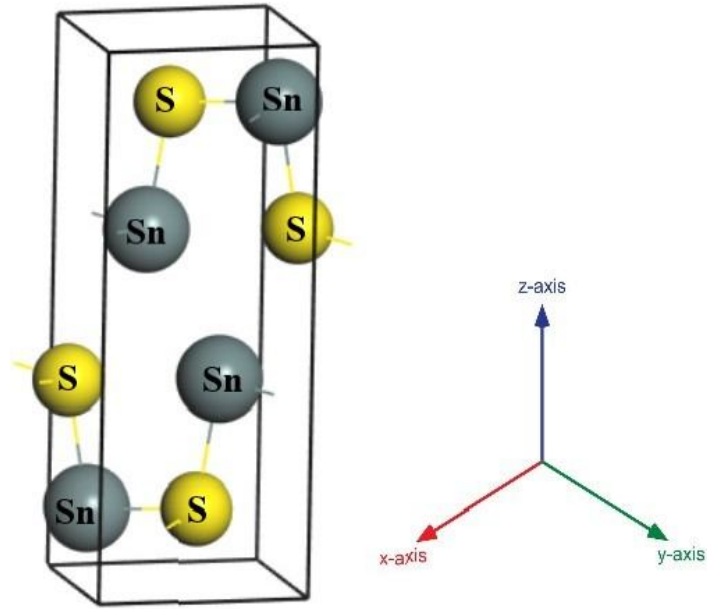

**Figure 2.** Three-dimensional schematic of the crystal structure of the SnS compound.

The total energy of the crystal is minimized by the process of geometry optimization at zero pressure and temperature via the use of CASTEP code using generalized gradient approximation (GGA) and local density approximation (LDA) to determine the equilibrium lattice parameters for the compound SnS. After geometry optimization, fully relaxed lattice parameters of SnS at zero pressure are obtained. The optimized lattice parameters along with available theoretical and experimental lattice parameters are recorded in **Table 2**. It is observed that when the functionals GGA, and PBEsol are used, the parameters and cell volume thus obtained are closer to the reference values than when LDA is used. For this reason, the functional GGA (PBEsol) is ultimately used to perform the geometry optimization for the SnS compound.

**Table 2.** Optimized lattice parameters ($a$, $b$, $c$ in Å, equilibrium volume $V$ in Å$^3$) of SnS as compared to available experimental and theoretical data.

| Compounds | $a$ | $b$ | $c$ | $V$ | Functionals | Ref. |
|---|---|---|---|---|---|---|
| SnS | 10.77 | 3.90 | 4.16 | 174.74 | LDA | This work |
|  | 11.23 | 4.05 | 4.20 | 192.09 | GGA-PBEsol | This work |
|  | 11.39 | 4.05 | 4.34 | 200.09 | _ | Exp. [39] |
|  | 11.20 | 3.98 | 4.33 | 193.35 | _ | Theo. [39] |

Since this study is about the effect of pressure on SnS, so the changes in the lattice parameters along with volume with pressure are recorded in **Table 3**. The pressure is increased step by step from 0 GPa to 20 GPa through an interval of 5 GPa and the effect of this gradual increasing pressure on the lattice parameters and cell volumes are recorded. It is seen from the table that there is a gradual decrease in the value of the lattice parameters as well as the cell volume as the pressure is gradually increased. This is to be expected since the increase in pressure acts to compress the crystal which in turn causes a reduction in the cell volume and corresponding lattice parameters [39,40].

**Table 3.** The effect of pressure on the optimized lattice parameters (*a*, *b*, and *c* in Å) and cell volume *V* in Å$^3$ of SnS.

| Compound | Pressure (GPa) | *a* | *b* | *c* | *V* | Ref. |
|---|---|---|---|---|---|---|
| SnS | 0 | 11.23 | 4.05 | 4.20 | 192.09 | This work |
| | 5 | 10.90 | 3.98 | 3.99 | 173.14 | This work |
| | 10 | 10.66 | 3.91 | 3.91 | 163.01 | This work |
| | 15 | 10.47 | 3.85 | 3.86 | 155.56 | This work |
| | 20 | 10.31 | 3.80 | 3.82 | 149.64 | This work |
| | - | 11.39 | 4.05 | 4.34 | 200.09 | Expt. [39] |
| | - | 11.20 | 3.98 | 4.33 | 193.35 | Theo. [39] |

From **Table 3**, it is observed that the equilibrium volume obtained from the experiment using GGA-PBEsol was slightly lower than the experimental values. This happened partly because of the reason that the experimental structural parameters were being studied at room temperature whereas the *ab initio* calculations were carried out assuming absolute zero temperature.

### 3.2 Elastic Properties

The link between the mechanical and dynamic behavior of the crystals under stress is provided by the elastic constants [41]. They provide important information regarding the response of the crystal to external forces which is characterized by the bulk modulus (*B*), shear modulus (*G*), Young's modulus (*Y*), and Poisson's ratio (*v*). These constants provide great assistance in determining the material strength [42,43] and determine its prospect in various engineering applications. Also, the calculation of the elastic constants is very important in ensuring the structural/mechanical stability of the systems. The calculated elastic constants are tabulated in **Table 4**. For the SnS system, the change in the elastic constants due to the application of pressure from 0 GPa to 20 GPa is also recorded in **Table 4**. The modified Born-Huang stability conditions for an orthorhombic crystal system are given below [44]:

$$C_{11} > 0, C_{44} > 0, C_{55} > 0, C_{66} > 0, C_{11}C_{22} > C^2_{12},$$

$$C_{11} C_{22} C_{33} + 2 C_{12} C_{13} C_{23} - C_{11} C^2_{23} - C_{22}C^2_{13} - C_{33}C^2_{12} > 0 \qquad (4)$$

**Table 4.** Single-crystal elastic constants, $C_{ij}$ for SnS at different pressures (all in GPa).

| Compound | Pressure | $C_{11}$ | $C_{22}$ | $C_{33}$ | $C_{44}$ | $C_{55}$ | $C_{66}$ | $C_{12}$ | $C_{13}$ | $C_{23}$ | Ref. |
|---|---|---|---|---|---|---|---|---|---|---|---|
| SnS | 0 | 73.91 | 90.55 | 47.95 | 49.69 | 24.07 | 25.66 | 20.73 | 26.70 | 40.84 | This work |
| | 5 | 125.34 | 124.94 | 129.84 | 86.74 | 33.74 | 34.42 | 39.65 | 39.51 | 78.07 | This work |
| | 10 | 149.84 | 141.18 | 160.73 | 110.85 | 34.98 | 41.36 | 58.09 | 48.84 | 100.96 | This work |
| | 15 | 194.36 | 165.85 | 191.80 | 130.48 | 44.52 | 47.91 | 71.57 | 65.91 | 126.87 | This work |
| | 20 | 234.72 | 191.68 | 210.92 | 148.76 | 47.48 | 53.16 | 83.59 | 76.92 | 150.80 | This work |
| | - | 83.89 | 99.34 | 52.42 | 51.40 | 27.93 | 28.35 | 24.42 | 29.48 | 42.71 | [39] |

From the data in **Table 4**, it is seen that none of the values are negative and the elastic constants for SnS satisfy the mechanical stability criteria. This suggested that the system SnS is mechanically stable. It is also seen that the ground state elastic constants obtained herein show good agreement with the previously computed values.

Among the nine independent single-crystal elastic constants of orthorhombic crystals, $C_{11}$, $C_{22}$ and $C_{33}$ are known to measure the response generated by the uniaxial strain and determined the resistance provided by the solid to the application of mechanical stress along the crystallographic *a*-, *b*- and *c*-directions respectively. It is seen from **Table 4** that at ambient condition $C_{33} < C_{11} < C_{22}$. This implies that as far as uniaxial strains are concerned, the crystal is most compressible when the applied stress is along the *c*-direction and it is least compressible when the stress is along the *b*-axis. This is a consequence of the layered structure of SnS. The situation changes drastically with the application of pressure. At 5 GPa, the magnitudes of $C_{11}$, $C_{22}$ and $C_{33}$ become almost identical. At 20 GPa $C_{11} > C_{33} > C_{22}$. Such non-monotonic pressure dependence of $C_{ii}$ implies that the bonding nature and strength are complex functions of pressure.

The elastic constant $C_{44}$ is used to represent the resistance offered by the compound to the shear deformation concerning tangential stress applied across the (001) plane in the [010] direction. If the elastic constants fulfilled the condition: $C_{44} < (C_{11}, C_{22}, C_{33})$ then it indicated that the shear along any of the three crystallographic directions can deform the compounds more easily than unidirectional compression along with those same directions. Since in all cases from **Table 4** it can be seen that this condition is satisfied, the compound which is being studied is more susceptible to shear deformations. The mechanical failure of SnS is thus expected to be controlled by shearing strain. Among the other elastic constants recorded, $C_{12}$ and $C_{13}$ are known as the off-diagonal shear components and are related to the resistance that the compound offers due to shape distortions of various natures. It is noticed that the ambient condition elastic constants calculated here agree quite well with those found in an earlier study [39]

From the single crystal elastic constants, $C_{ij}$'s different elastic moduli, anisotropy indicators, and Poisson's ratio can be obtained. According to the Voigt's approximation [45], from the linear combination of various elastic constants, the isotropic bulk and shear modulus can be calculated. The Voigt approximated bulk and shear moduli are indicated by $B_V$ and $G_V$, respectively. On the other hand, Ruess [46] derives a different estimate for isotropic bulk and shear moduli from the single crystal elastic constants. The Ruess approximated bulk and shear moduli are indicated by $B_R$ and $G_R$, respectively. Later Hill [47] proved that the Voigt-approximated values give the upper limit of the polycrystalline elastic moduli whereas the Ruess-approximated values give the lower limit of the polycrystalline elastic moduli. The real value lies between the Voigt and Reuss bounds. Hill proposed that the arithmetic mean given by $B = (B_V+B_R)/2$ for the bulk modulus and $G = (G_V+G_R)/2$ for the shear modulus are closer to the real situation. The polycrystalline bulk modulus ($B$), shear modulus ($G$), Pugh's ratio ($B/G$), Young's modulus ($Y$), and Poisson's ratio ($v$) are recorded in **Table 5**. Since this investigation mainly focuses on the study of the effect of pressure on different properties of the compound SnS, the change in the bulk, shear and Young's modulus, Pugh's and Poisson's ratios for pressures from 0 GPa to 20 GPa are also recorded in **Table 5** given below.

**Table 5.** The bulk, shear, and Young's modulus (all in GPa), Pugh's and Poisson's ratio for the compound SnS together with the values obtained at different pressures (all in GPa).

| Compound | Pressure | $B_V$ | $B_R$ | $B$ | $G_V$ | $G_R$ | $G$ | B/G | Y | v | Ref. |
|---|---|---|---|---|---|---|---|---|---|---|---|
| SnS | 0 | 43.22 | 40.55 | 41.88 | 28.16 | 21.42 | 24.79 | 1.69 | 62.12 | 0.25 | This work |
| | 5 | 77.17 | 75.93 | 76.55 | 45.84 | 38.09 | 41.96 | 1.82 | 106.44 | 0.27 | This work |
| | 10 | 96.39 | 94.81 | 95.60 | 53.69 | 41.28 | 47.49 | 2.01 | 122.23 | 0.29 | This work |
| | 15 | 120.08 | 118.82 | 119.45 | 63.76 | 47.69 | 55.73 | 2.14 | 144.68 | 0.30 | This work |
| | 20 | 139.10 | 139.27 | 139.63 | 71.61 | 50.60 | 61.10 | 2.29 | 159.99 | 0.30 | This work |
| | - | 47.65 | 44.57 | 46.11 | 30.80 | 24.48 | 27.68 | 1.67 | 69.12 | 0.25 | [39] |

When tension or compression is applied to a material along its length then the ability of the material to resist this change is measured by the Young's modulus and can also be obtained by dividing tensile stress by tensile strain. It is also found that as Young's modulus increases the covalent nature of the compounds increases along with it [48]. From the values given in **Table 5**, it can be seen that with pressure the value of Young's modulus increases. The Pugh's ratio, which is the ratio between bulk modulus, $B$, and shear modulus, $G$ (i.e., $B/G$), indicates the ductility or the brittleness of a material. It is known that, if the value of the Pugh's ratio exceeds the value of 1.75, then the material will exhibit ductile behavior whereas if the value of the Pugh's ratio is less than 1.75, then the material would be brittle in nature. From the tabulated values in **Table 5**, it can be seen that at 0 GPa, SnS has a Pugh's ratio which is less than 1.75, and therefore, at 0 GPa SnS is brittle in nature. But it is found that, with the application of pressure, the value of Pugh's ratio for SnS becomes greater than 1.75. That is, the application of pressure causes SnS to exhibit ductile behavior. The threshold between ductility and brittleness in a material can be indicated by another quantity which is known as the Poisson's ratio. The Poisson's ratio has a critical value of 0.26. If the value of Poisson's ratio exceeds this critical value then the material is expected to exhibit ductile behavior, but if the value is less than this critical value, then the material should be brittle in nature. A relationship has been observed between the Poisson's ratio and the nature of interatomic forces in solids [49]. It is also known that, if the value of the Poisson's ratio resides within the range 0.25 to 0.50 then the central force interactions dominate the bondings in the solid. But if the value of the ratio is outside this given range, then the solid is dominated by the non-central force interactions. Furthermore, whether ionic and covalent bonding is present within a compound can also be indicated by the Poisson's ratio. For instance, in ionic and covalent materials, the Poisson's ratio is observed to have values of 0.25 and 0.10, respectively [50]. From **Table 5**, since the compound with and without the effect of pressure shows a value of Poisson's ratio which is within the given range, it could be said that the bonding in the compound is dominated by the central forces. Also, as seen in the case of Pugh's ratio, since the value of Poisson's ratio for SnS at 0 GPa does not exceed the critical value, it should exhibit brittle behavior. For all the other pressures, since SnS shows Poisson's ratio greater than the critical value, they are expected to be ductile in nature.

Some more other important mechanical performance indicators, namely the machinability index ($\mu^M$), Kleinman parameter ($\zeta$), and Vickers hardness ($H_V$) are calculated using the following widely employed formulae [51,52] and are listed in **Tables 6 and 7.**

$$\mu^M = \frac{B}{C_{44}} \qquad (5)$$

$$\zeta = \frac{C_{11} + 8C_{12}}{7C_{11} + 2C_{12}} \qquad (6)$$

$$H_V = \frac{(1-2\nu)Y}{6(1+\nu)} \qquad (7)$$

To determine the dry lubricating nature of a material, a quantity called the machinability index ($\mu^M$) is used. The quality of lubrication is directly proportional to the value of $\mu^M$, i.e., the higher the value the greater the lubricating nature of the material. Also, a higher value of $\mu^M$ indicates that it has a good level of machinability [53-57]. Generally, the value of the Kleinman parameter ($\zeta$) is observed to lie between 0 and 1. According to Kleinman, the significant contribution to the resistance offered by bond stretching or contracting towards external stress is represented by the lower limit of $\zeta$, whereas the significant contribution to the resistance offered by bond bending towards external stress is represented by the upper limit of $\zeta$. To obtain a clear understanding of the elastic and plastic properties of a material, information regarding the hardness is needed. In **Table 6,** the Vickers hardnesses at different pressures are listed.

**Table 6.** Calculated Vickers hardness, $H_V$, and the effect of pressure on it for the SnS compound.

| Compound | Pressure (GPa) | Vickers Hardness, $H_V$ (GPa) |
|---|---|---|
| SnS | 0 | 4.141 |
| | 5 | 6.426 |
| | 10 | 6.633 |
| | 15 | 7.419 |
| | 20 | 8.205 |

When pressure is applied to the compound, Vickers Hardness of the compound increases. The increment is non-monotonic. The increment is very small in the pressure range from 5 to 10 GPa.

The directional bulk modulus, which are represented by $B_a$, $B_b$ and $B_c$ along *a*-, *b*- and *c*-axis, respectively, and the isotropic bulk modulus ($B_{relax}$), are defined as follows [52,53]:

$$B_a = a\frac{dP}{da} = \frac{\Lambda}{1 + \alpha + \beta} \tag{8}$$

$$B_b = b\frac{dP}{db} = \frac{B_a}{\alpha} \tag{9}$$

$$B_c = c\frac{dP}{dc} = \frac{B_a}{\beta} \tag{10}$$

$$B_{relax} = \frac{\Lambda}{(1 + \alpha + \beta)^2} \tag{11}$$

where,

$$\Lambda = C_{11} + 2C_{12}\alpha + C_{22}\alpha^2 + 2C_{13}\beta + C_{33}\beta^2 + 2C_{23}\alpha\beta$$

and,

$$\alpha = \frac{\{(C_{11} - C_{12})(C_{33} - C_{13})\} - \{(C_{23} - C_{13})(C_{11} - C_{13})\}}{\{(C_{33} - C_{13})(C_{22} - C_{12})\} - \{(C_{13} - C_{23})(C_{12} - C_{23})\}}$$

$$\beta = \frac{\{(C_{22} - C_{12})(C_{11} - C_{13})\} - \{(C_{11} - C_{12})(C_{23} - C_{12})\}}{\{(C_{22} - C_{12})(C_{33} - C_{13})\} - \{(C_{12} - C_{23})(C_{13} - C_{23})\}}$$

The relevant parameters obtained from these relations for the SnS system are also recorded in **Table 7**. We have only shown these values for the crystal structure in the ground state in **Table 7** as the pressure variation is weak.

**Table 7.** The machinability index ($\mu^M$), Kleinman parameter ($\zeta$), Vickers hardness ($H_V$ in GPa), bulk modulus ($B_{relax}$ in GPa), bulk modulus along *a*-, *b*- and *c*-axis ($B_a$, $B_b$, $B_c$ in GPa), $\alpha$ and $\beta$ for the compound SnS.

| Compound | $\mu^M$ | $\zeta$ | $B_{relax}$ | $B_a$ | $B_b$ | $B_c$ | $\alpha$ | $\beta$ |
|---|---|---|---|---|---|---|---|---|
| SnS | 0.843 | 0.429 | 40.143 | 130.063 | 337.826 | 70.077 | 0.386 | 1.856 |

From **Table 7**, it is observed that the machinability index of SnS is low and this compound has low dry lubricity. The Kleinman parameter lies in between 0 and 1, thus both bond stretching and bond bending contributions play role in determining the mechanical strength of SnS.

The dependence of the mechanical properties on different crystallographic directions is studied with the help of the elastic anisotropy indices [57-59,60]. In the design of materials, especially layered compounds, the study of elastic anisotropy is extremely helpful. The elastic anisotropy indices of SnS are investigated and the Zener anisotropy factor ($A$), shear anisotropy factors ($A_1$, $A_2$, $A_3$), percentage anisotropy in compressibility ($A_B$) and shear ($A_G$), universal anisotropy factor ($A^U$, $d_E$), equivalent Zener anisotropy factor ($A^{eq}$) are calculated and recorded in **Table 8**. The anisotropy factors were calculated from the following relations which are widely employed [61-63]:

$$A = \frac{2C_{44}}{C_{11} - C_{12}} \quad (12)$$

$$A_1 = \frac{4C_{44}}{C_{11} + C_{33} - 2C_{13}} \quad (13)$$

$$A_2 = \frac{4C_{55}}{C_{22} + C_{33} - 2C_{23}} \quad (14)$$

$$A_3 = \frac{4C_{66}}{C_{11} + C_{22} - 2C_{12}} \quad (15)$$

$$A_B = \frac{B_V - B_R}{B_V + B_R} \quad (16)$$

$$A_G = \frac{G_V - G_R}{G_V + G_R} \quad (17)$$

$$A^U = \frac{B_V}{B_R} + 5\frac{G_V}{G_R} - 6 \geq 0 \tag{18}$$

$$d_E = \sqrt{A^U + 6} \tag{19}$$

$$A^{eq} = \left(1 + \frac{5}{12}A^U\right) + \sqrt{\left(1 + \frac{5}{12}A^U\right)^2 - 1} \tag{20}$$

The value of unity of the calculated $A$, $A_1$, and $A_2$ indicates the isotropic nature of the material. From **Table 8**, it can be seen that the calculated values of $A$, $A_1$, and $A_2$ are found to be different from the unity which represents that the compound SnS is strongly anisotropic for shearing stress in different crystal planes. In the cases of $A_B$ and $A_G$, the elastic isotropy is represented by the value zero whereas the highest anisotropy is represented by the value 1. The universal anisotropic factor denoted by $A^U$, applies to all kinds of crystal symmetries and has been proposed by Ranganathan and Ostoja-Starzewski [64]. When $A^U = 0$, the crystal is considered to be elastically isotropic. Any other value of $A^U$ other than zero indicates the anisotropic nature of the crystal. Similarly, in the case of the quantity $A^{eq}$, the value of 1 represents an isotropic crystal while any value other than 1 represents the anisotropic behavior of a crystal. Thus, from **Table 8** it can be concluded that the compound SnS is highly anisotropic. All the anisotropy indices reflected structural/mechanical anisotropy that originated from the anisotropy in the bonding strengths in different directions in the unit cell of the SnS crystal.

The universal log-Euclidean index is defined by the following equation [62]:

$$A^L = \sqrt{\left[ln\left(\frac{B_V}{B_R}\right)\right]^2 + 5\left[ln\left(\frac{C_{44}^V}{C_{44}^R}\right)\right]^2} \tag{21}$$

where,

$$C_{44}^R = \frac{5}{3}\frac{C_{44}(C_{11} - C_{12})}{3(C_{11} - C_{12}) + 4C_{44}} \quad \text{is the Reuss value of } C_{44}$$

and,

$$C_{44}^V = C_{44}^R + \frac{3}{5}\frac{(C_{11} - C_{12} - 2C_{44})^2}{3(C_{11} - C_{12}) + 4C_{44}} \quad \text{is the Voigt value of } C_{44}$$

According to Kube and Jong [65], the inorganic crystalline compounds had the value of $A^L$ which lies in the range $0 \leq A^L \leq 10.26$, and 90% of these compounds have $A^L < 1$. In the case of a perfect isotropic crystal, it is observed that $A^L = 0$. Although it is difficult to ascertain whether a solid is layered or not depending solely on the value of $A^L$, the majority (78%) of these inorganic crystalline compounds with high $A^L$ value exhibit layered structural features [65]. The high anisotropy and layered nature of the titled compounds can be predicted from the above discussions. Besides, there are measures of anisotropy in the uniaxial bulk modulus expressed as:

$$A_{B_a} = \frac{B_a}{B_b} = \alpha \tag{22}$$

$$A_{B_c} = \frac{B_c}{B_b} = \frac{\alpha}{\beta} \tag{23}$$

All of the above-mentioned parameters were calculated and their values were recorded in **Table 8** as shown below.

**Table 8.** Elastic anisotropy indices of the compound SnS in the ground state.

| Compound | $A$ | $A_1$ | $A_2$ | $A_3$ | $A_B$ | $A_G$ | $A^U$ | $d_E$ | $A^{eq}$ | $A^L$ | $A_{B_a}$ | $A_{B_c}$ |
|---|---|---|---|---|---|---|---|---|---|---|---|---|
| SnS | 1.87 | 2.90 | 1.69 | 0.83 | 0.03 | 0.14 | 1.64 | 2.76 | 3.04 | 0.57 | 0.36 | 0.21 |

All the results displayed in **Tables 6-8** are novel and cannot be compared with any other work at this moment.

It is expected that three-dimensional (3D) direction-dependent Young's modulus, shear modulus, compressibility (inverse of the bulk modulus), and Poisson's ratio for an isotropic solid are expected to exhibit spherical shapes. If they deviate from the spherical shapes it indicates the presence of anisotropy. The ELATE [66] generated 3D plots of directional dependence of Young's modulus, shear modulus, compressibility, and Poisson's ratio for the SnS system are illustrated in the **Figures 3** below,

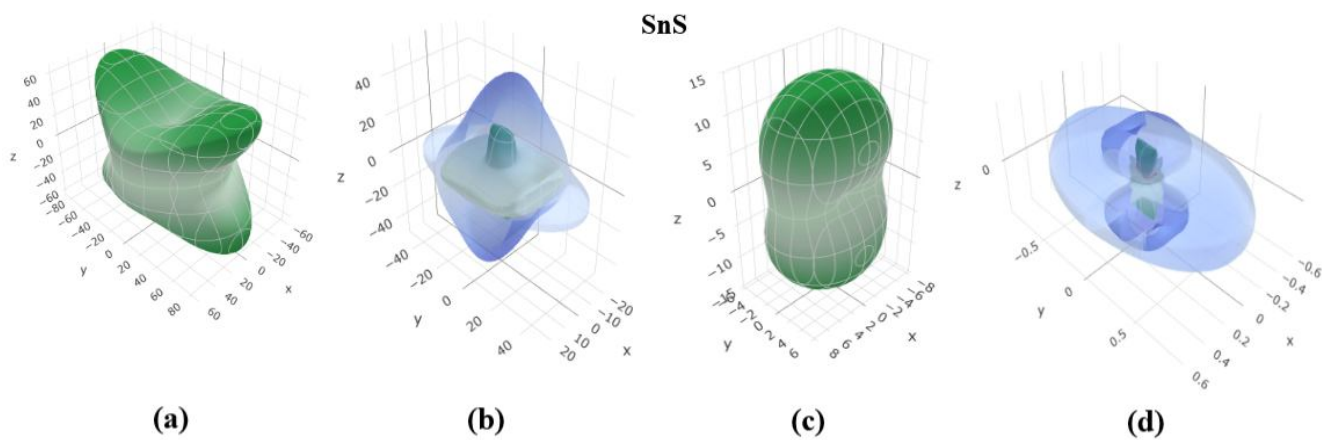

**Figure 3.** 3D directional dependences of (a) Young modulus (b) shear modulus (c) compressibility and (d) Poisson's ratio for the compound SnS.

The 3D plots clearly demonstrate the anisotropic nature of the elastic parameters of SnS. The anisotropy is particularly prominent in the out of the *ab*-plane direction.

### 3.3 (a) Electronic Band Structure

The electronic band structure illustrates the variation of electron energy as a function of momentum along different directions within the Brillouin zone. The diagrams of the band structure of the optimized SnS system at various pressures are presented in **Figure 4**. In the figure, *Γ*, *Z*, *T*, *Y*, *S*, *X*, *U*, *R* designate the (0 0 0), (0 0 ½), (-½ 0 ½), (-½ 0 0), (-½ ½ 0), (0 ½ 0), (0 ½ ½), (-½ ½ ½) high symmetry points, respectively. The direct and indirect band gap values of the SnS compound at different pressures are given in **Table 9**.

**Table 9.** Calculated values of band gaps at different pressures of the SnS compound.

| Compound | Pressure (GPa) | Band gap ($E_g$) type | Band gap ($E_g$) (eV) |
|---|---|---|---|
| SnS | 0 | Indirect | 0.635 |
|  |  | Direct | 0.592 |
|  | 5 | Indirect | 0 |
|  |  | Direct | 0 |
|  | 10 | Indirect | 0 |
|  |  | Direct | 0 |
|  | 15 | Indirect | 0 |
|  |  | Direct | 0 |
|  | 20 | Indirect | 0 |
|  |  | Direct | 0 |

(a)

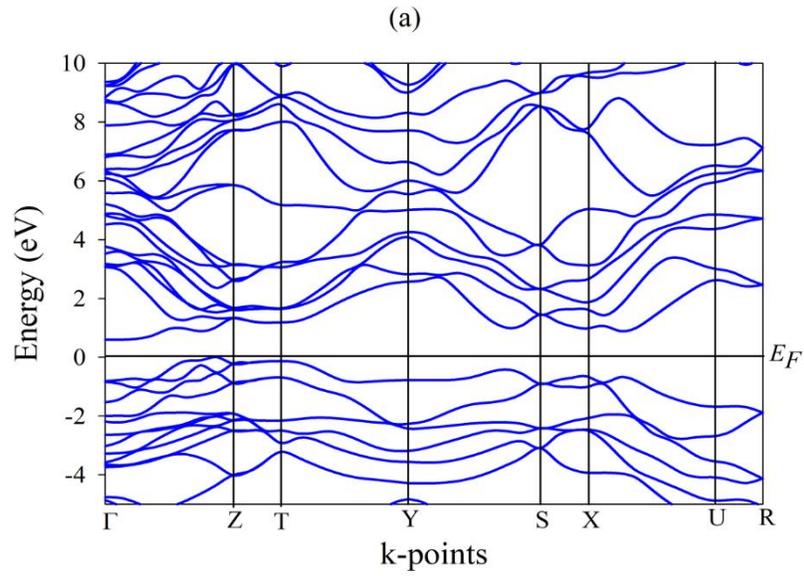

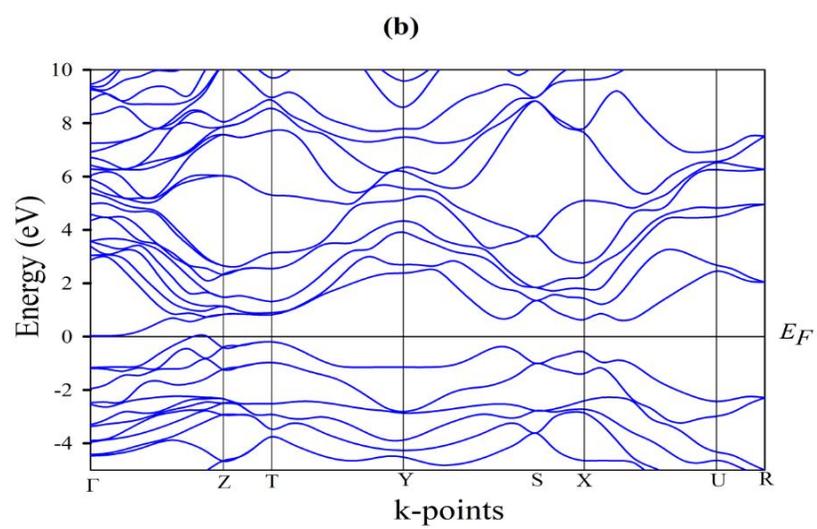

(b)

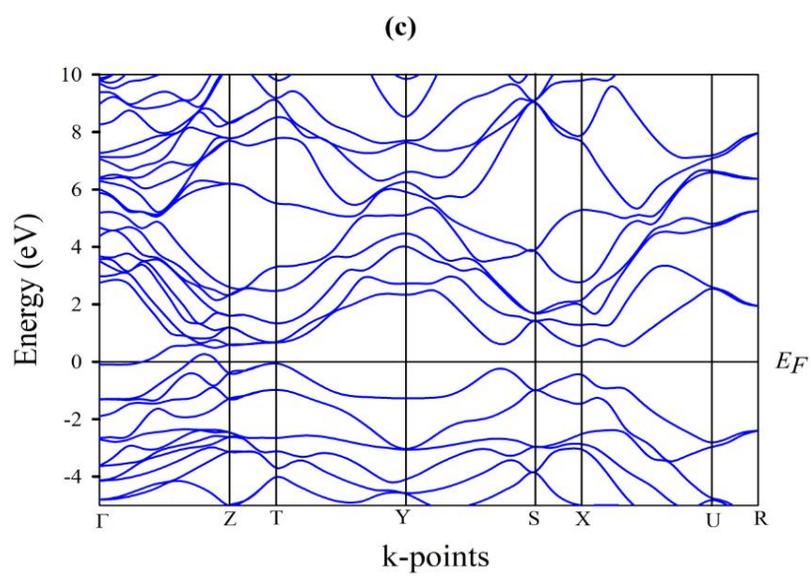

(c)

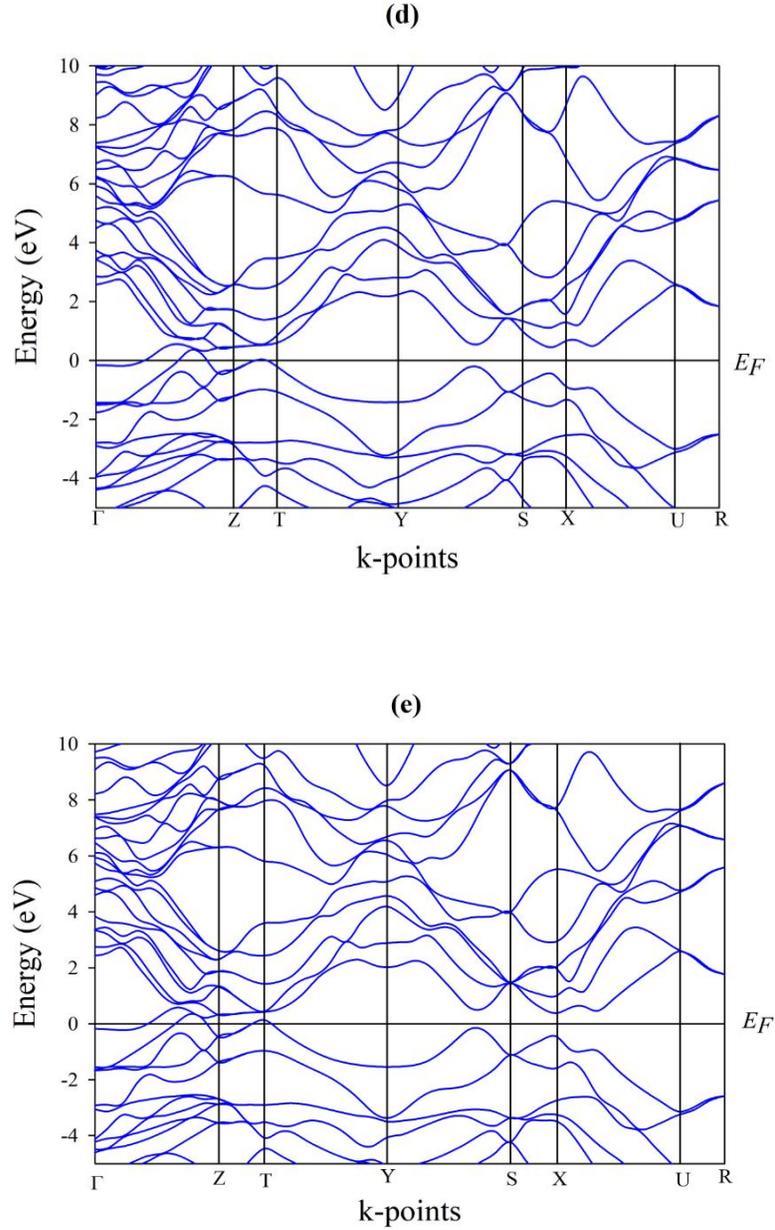

**Figure 4.** The band structure of SnS along the high symmetry directions of the k-space within the first Brillouin zone at (a) 0 GPa, (b) 5 GPa, (c) 10 GPa, (d) 15 GPa (e) 20 GPa pressures.

From the panels shown in **Figure 4**, it can be seen that SnS is a narrow gap semiconductor in the ground state. But immediately after the application of 5 GPa pressure, it becomes a conductor as the electronic dispersion curve (EDC) running along the $\Gamma$-Z direction crosses weakly the Fermi level (semi-metallic behavior). As the pressure is increased further, its semi-metallic properties gradually increase and at 20 GPa several bands cross the Fermi level, showing completely metallic features. From the EDCs of SnS, it can be seen that the electronic band dispersion is greater in the regions from *Γ* to *Z*, from *T* to *Y*, from *Y* to *S*, and from *X* to *U*. Because of this, the electrons in these regions have lower effective mass and greater mobility. The band dispersion is relatively less in the regions from *Z* to *T*, from *S* to *X*, and from *U* to *R*. Because of this, the electrons in these regions have relatively greater effective mass and their mobility is comparatively lower. Almost all the bands close to the Fermi energy are dispersive. Absence of flat bands indicates that the electronic energy density of states close to the Fermi energy should be low [67] in the metallic state of SnS.

**(b) Electronic Density of States (EDOS)**

The number of permitted energy states that can be occupied by the electrons per unit of energy interval is described by the electronic density of states (EDOS). The total density of states (TDOS) and the atomic orbital specific partial density of states (PDOS) of the SnS compound are calculated and the effect of pressure on TDOS and PDOS for the system is studied by gradually increasing pressure from 0 GPa to 20 GPa. These results are illustrated in **Figure 5(a-e)** as a function of energy, ($E$-$E_F$). The Fermi level is represented by the vertical dashed lines placed at 0 eV. From **Figure 5**, it can be seen that TDOS for SnS is zero at the Fermi level at 0 GPa. Therefore, SnS is a semiconducting material in the ground state. But as pressure is increased from 5 GPa to 20 GPa, a gradual increase in the value of TDOS at the Fermi surface is obtained which indicates the increasing metallic nature of the SnS material with the increasing pressure in agreement with the band structure results. For SnS, from **Figure 5**, it is seen that, in the case of the valence band in the lower energy from -5.0 eV to 0 eV, the contribution of S-3*p*, Sn-5*s*, and Sn-5*p* electronic states are significant. It is also observed from **Figure 5** that the contribution of the S-3*s* orbital to the valence band in the case of the SnS system is minimal. For this system, it is seen that as pressure is increased from 0 GPa to 20 GPa, the bands due to S-3*p*, Sn-5*s*, and Sn-5*p* electrons crossed the Fermi level which results in the gradual increase in metallic behavior. For the conduction band near the Fermi level, in the case of SnS, the dominant contribution comes from the Sn-5*p* orbital. The bonding peak in the electronic energy density of states is defined as the nearest peak at the negative energy below the Fermi level in TDOS whereas the anti-bonding peak is defined as the nearest peak at the positive energy above the Fermi level in TDOS. The difference in energy between the bonding peak and the anti-bonding peak is defined as the pseudo gap. A pseudogap very close to $E_F$ is an indication of high structural stability [68-71]. For the SnS system, the bonding and anti-bonding peaks are located within 2 eV of the Fermi level. The close proximity of the peaks in the TDOS to the Fermi energy indicates that it is possible to change the electronic properties of SnS by suitable atomic substitution (alloying) or by applying moderate pressure.

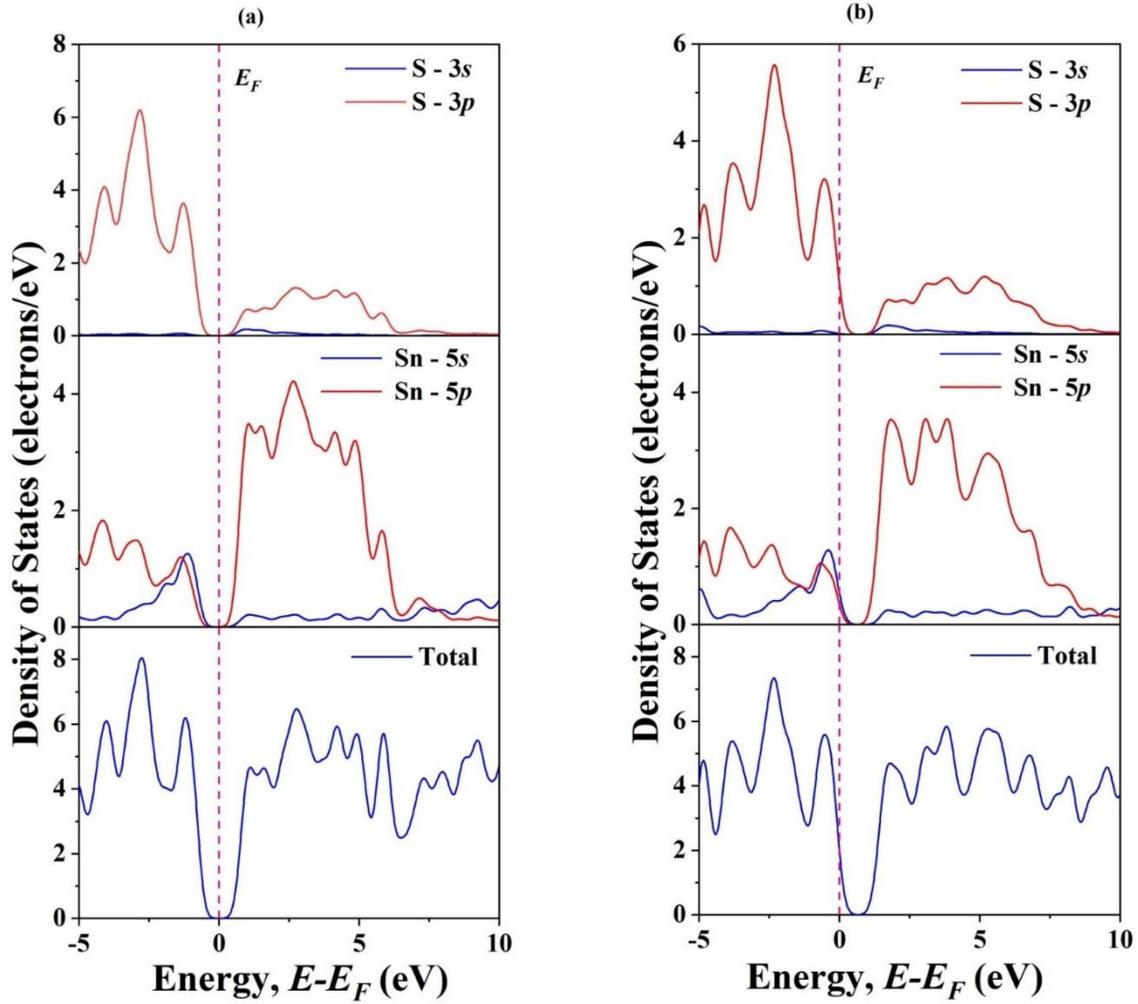

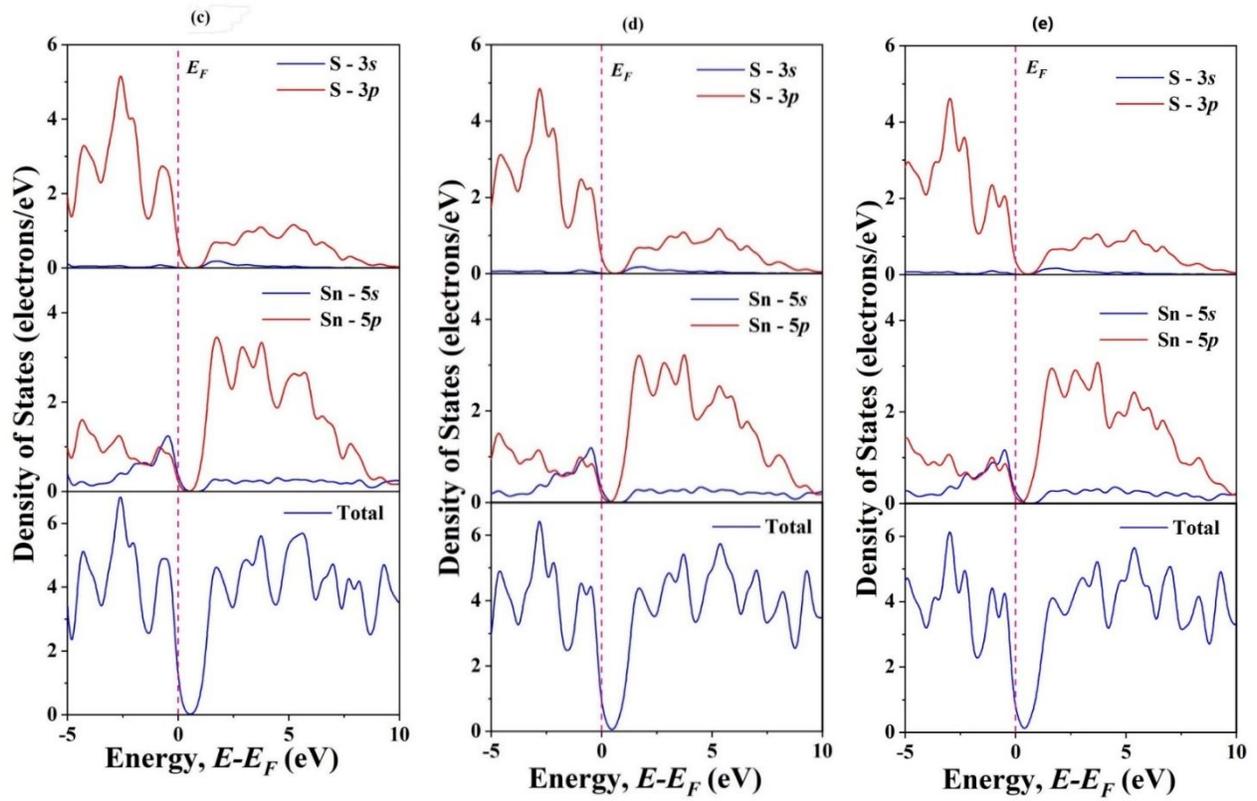

**Figure 5.** Total and partial electronic density of states of SnS at (a) 0 GPa, (b) 5 GPa (c) 10 GPa, (d) 15 GPa, and (e) 20 GPa pressures.

In **Figure 6** the total density of states of the SnS compound are illustrated at different pressures together to compare the pressure induced variation. With increasing pressure the Fermi level shifts towards a bonding peak in the TDOS. This is an indication of pressure induced structural instability.

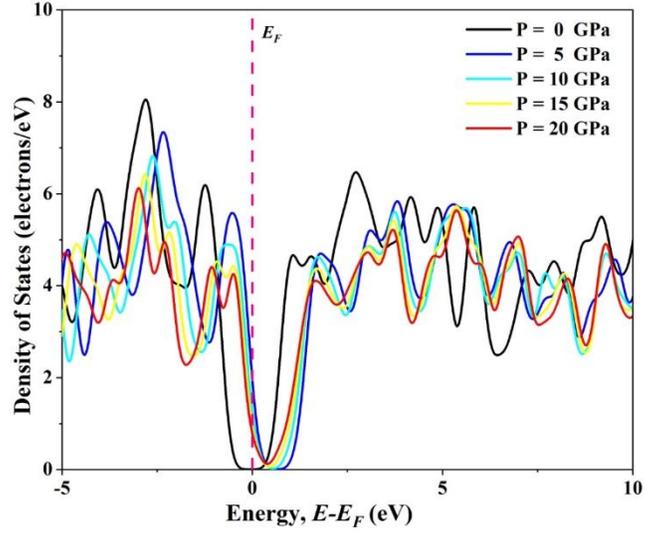

**Figure 6.** Comparison of the total density of states of SnS at 0 GPa, 5 GPa, 10 GPa, 15 GPa, and 20 GPa pressures.

**(c) Coulomb Pseudopotential**

The following relation has been used to calculate the electron-electron Coulomb interaction parameter [72]:

$$\mu^* = \frac{0.26 N(E_F)}{1 + N(E_F)} \quad (24)$$

This electron-electron interaction parameter is also known as the Coulomb pseudopotential. This is an interesting parameter which measures the repulsive Coulomb correlations in solids. The repulsive Coulomb pseudopotential inhibits Cooper pairing in superconducting systems and reduces the superconducting critical temperature [73]. The computed values of $\mu^*$ of SnS at different pressures are presented in the **Table 10** below.

**Table 10.** Calculated values of the Coulomb pseudopotential, $\mu^*$ of SnS for different values of pressure.

| Compound | Pressure (GPa) | Coulomb pseudopotential $\mu^*$ |
|---|---|---|
| SnS | 0 | --- |
| | 5 | 0.174 |
| | 10 | 0.147 |
| | 15 | 0.128 |
| | 20 | 0.116 |

It is instructive to note that the Coulomb pseudopotential decreases with increasing pressure. This is an indication that the electronic correlations weaken as the compound becomes more and more metallic. Superconductivity in SnS emerges in the highly metallic state at higher pressures with very small Coulombic repulsion parameter.

### 3.3 Thermo-physical Properties

**(a) Debye Temperature**

The highest possible frequency of a normal mode of vibration of an atom in a crystal is defined as the Debye frequency. Equivalently it can be represented by the Debye temperature, $\theta_D$. This particular parameter helps to correlate the elastic properties with the thermodynamic properties such as phonons, thermal expansion, thermal conductivity, specific heat, and lattice enthalpy of a solid. It is also related to the bonding strength among the atoms present in the compounds. The characteristic energy scale for the phonons involved in the Cooper pairings is also set by $\theta_D$ in the conventional superconductors. Mean sound velocity, $v_m$, and mass density, $\rho$ of a solid can be used to calculate the Debye temperature. With the help of the expressions [74] given below, the parameters calculated were recorded in **Table 11**.

$$\theta_D = \frac{h}{k_B}\left[\left(\frac{3n}{4\pi}\right)\frac{N_A\rho}{M}\right]^{\frac{1}{3}} v_m \qquad (25)$$

In (25), $h$ represents the Planck's constant, $k_B$ represents the Boltzmann's constant, $N_A$ denotes the Avogadro's number, $\rho$ denotes mass density, $M$ denotes the molecular weight and $n$ represents the number of atoms in the cell.

The mean sound velocity, $v_m$ in the crystal can be calculated from:

$$v_m = \left[\frac{1}{3}\left(\frac{1}{v_l^3} + \frac{2}{v_t^3}\right)\right]^{-\frac{1}{3}} \qquad (26)$$

where the longitudinal and transverse sound velocities were represented by $v_l$ and $v_t$, respectively. The longitudinal and transverse sound velocities can be obtained in terms of bulk modulus, $B$ and shear modulus $G$ from the following expressions:

$$v_l = \left[(B + \frac{4G}{3})/\rho\right]^{\frac{1}{2}} \qquad (27)$$

and,

$$v_t = \left[\frac{G}{\rho}\right]^{\frac{1}{2}} \qquad (27)$$

**Table 11**. Calculated mass density ($\rho$ in gm cm$^{-3}$), longitudinal, transverse and mean sound velocities ($v_l$, $v_t$, and $v_m$ in km sec$^{-1}$), and Debye temperature ($\theta_D$ in K) of SnS at different pressures.

| Compound | Pressure (GPa) | $\rho$ | $v_l$ | $v_t$ | $v_m$ | $\theta_D$ |
|---|---|---|---|---|---|---|
| SnS | 0 | 5.27 | 3770.80 | 2168.85 | 2408.63 | 249.74 |
|  | 5 | 5.79 | 4783.59 | 2691.99 | 2995.12 | 320.40 |
|  | 10 | 6.14 | 5229.19 | 2956.35 | 3288.0 | 358.78 |
|  | 15 | 6.44 | 5644.67 | 3147.33 | 3504.35 | 363.29 |
|  | 20 | 6.69 | 5903.75 | 3270.64 | 3644.79 | 409.22 |

**(b) Heat Capacity**

Another important intrinsic thermodynamic property of a material is the heat capacity. A high value of heat capacity in a material indicates a high value of thermal conductivity and a low value of thermal diffusivity. The change in thermal energy per unit volume in a material per degree Kelvin change in the temperature is defined by the heat capacity per unit volume ($\rho C_P$). The heat capacity per unit volume ($\rho C_P$) of a material can be expressed as [75,76]:

$$\rho C_P = \frac{3k_B}{\Omega} \qquad (29)$$

where $N = 1/\Omega$ represents the number of atoms per unit volume. The heat capacity per unit volume of SnS is recorded in **Table 12**.

**(c) Melting Temperature**

To obtain the melting temperature of solids via the elastic constants, an empirical formula was put forward by Fine *et al.* [77]:

$$T_m = 354 + 1.5(2C_{11} + C_{33}) \qquad (30)$$

This expression is used to estimate the melting temperature of SnS roughly and it is recorded in **Table 12**. Melting temperature reflects the cohesive energy, strength of atomic bonding and lattice anharmonicity in crystalline solids. The computed melting temperature of SnS is quite low, indicating its relatively soft nature.

**Table 12.** Calculated heat capacity per unit volume ($\rho C_\rho$ in JK$^{-1}$m$^{-3}$) and melting temperature ($T_m$ in K) of the SnS compound in the ground state.

| Compound | $\rho C_\rho$ (×10$^6$) | $T_m$ |
|---|---|---|
| SnS | 1.74 | 647.67 |

**(d) Minimum Thermal Conductivity**

In the heated state or during cooling, the minimum thermal conductivity provides an understanding of the behavior of atoms inside a crystal. Minimum thermal conductivity is defined as the hypothetical lowest value of the inherent thermal conductivity. The minimum thermal conductivity could be calculated from the expression below [78-80]:

$$k_{min} = k_B v_m \left(\frac{nN_A\rho}{M}\right)^{\frac{2}{3}} \qquad (31)$$

The calculated values $k_{min}$ for SnS for pressures from 0 GPa to 20 GPa are recorded in **Table 13**.

**Table 13.** Calculated minimum thermal conductivity, K$_{min}$ of SnS for different pressures.

| Compound | Pressure (GPa) | Minimum Thermal Conductivity, K$_{min}$ (Wm$^{-1}$K$^{-1}$) |
|---|---|---|
| SnS | 0 | 0.3789 |
| | 5 | 0.4711 |
| | 10 | 0.5171 |
| | 15 | 0.5512 |
| | 20 | 0.5733 |

From the above **Table 13,** we can see that with increasing pressure, the minimum thermal conductivity of SnS also increases. This observation is consistent with the increase in the Debye temperature with increasing pressure.

**(e) Anisotropies in Sound Velocity**

It is observed that the velocity with which the sound waves (both longitudinal and transverse waves) travel through a solid is free from the frequency and the dimension of the material. The sound velocity inside the solid depends only on the nature of the material with which the solid is made of. Each atom in a solid consisted of only three modes of vibrations - one longitudinal mode and two transverse modes. The pure longitudinal and transverse modes of vibrations are obtained only along with certain crystallographic directions in the anisotropic crystals whereas the quasi-transverse or quasi longitudinal modes are found in all other directions. Since SnS is an orthorhombic system, the acoustic velocities along [100], [010], and [001] directions are expected to be different and can be computed from the expressions given below [81]:

$$[100]v_l = \sqrt{\frac{C_{11}}{\rho}} \quad , \quad [010]v_{t1} = \sqrt{\frac{C_{66}}{\rho}} \quad , \quad [001]v_{t2} = \sqrt{\frac{C_{55}}{\rho}} \quad (32)$$

$$[010]v_l = \sqrt{\frac{C_{22}}{\rho}} \quad , \quad [100]v_{t1} = \sqrt{\frac{C_{66}}{\rho}} \quad , \quad [001]v_{t2} = \sqrt{\frac{C_{44}}{\rho}} \quad (33)$$

$$[001]v_l = \sqrt{\frac{C_{33}}{\rho}} \quad , \quad [100]v_{t1} = \sqrt{\frac{C_{55}}{\rho}} \quad , \quad [010]v_{t2} = \sqrt{\frac{C_{44}}{\rho}} \qquad (34)$$

The calculated sound velocities along different crystallographic axes of SnS are presented in **Table 14**.

**Table 14.** Anisotropic sound velocities in SnS along different crystallographic directions (in ms$^{-1}$).

| Propagation directions | Sound velocity |
|---|---|
| $[100]v_l$ | 3744.89 |
| $[010]v_{t1}$ | 2206.62 |
| $[001]v_{t2}$ | 2137.10 |
| $[010]v_l$ | 4145.09 |
| $[100]v_{t1}$ | 2206.62 |
| $[001]v_{t2}$ | 3070.66 |
| $[001]v_l$ | 3016.44 |
| $[100]v_{t1}$ | 2137.10 |
| $[010]v_{t2}$ | 3070.66 |

**3.4 Bond Population Analysis**

The bonding nature of the SnS compound has been analyzed in detail through the investigations of the Mulliken population analysis (MPA) [36] and Hirshfeld population analysis (HPA) [82]. The results obtained are recorded in **Tables 15** and **Table 16**, respectively. According to the MPA, in SnS, the Sn atom is seen to give up 0.54 electronic charges to the S atom. As a result of the presence of partial ionic bonding, the transfer of electrons occurred between different atoms in the compound. The effective valence charge (EVC) is defined as the difference between formal ionic charge and calculated Mulliken charge. The presence of non-zero EVC atoms indicates that, in the compound, covalent bonding is also present among the atoms (**Table 15**). On the other hand, according to HPA, the Sn atom transfers 1.46 electronic charges to the S atom. The ionic bonding is stronger in the HPA. Because of being free from basis set dependence, the HPA results are considered to be overall more reliable. Compared to the HPA,

MPA leads to large changes in the computed atomic charges for small variations in the underlying basis sets and could overestimate the covalent character of a bond.

**Table 15.** Charge spilling parameter (%), orbital charges (electron), atomic Mulliken charge (electron), Hirshfeld charge (electron), and EVC (electron) of SnS compound.

| Compound | Atoms | s | p | Total | Mulliken charge | Hirshfeld charge | Charge spilling (%) | EVC |
|---|---|---|---|---|---|---|---|---|
| SnS | S | 1.90 | 4.64 | 6.54 | -0.54 | -0.20 | 1.94 | -1.46 |
|  | Sn | 1.81 | 1.65 | 3.46 | 0.54 | 0.20 |  | 1.46 |

**Table 16.** Calculated bond overlap population and bond lengths (Å) for the SnS compound.

| Compound | Bond | Population | Length |
|---|---|---|---|
| SnS | S 2 -- Sn 3 | 0.12 | 2.66494 |
|  | S 3 -- Sn 2 | 0.12 | 2.66494 |
|  | S 1 -- Sn 4 | 0.12 | 2.66494 |
|  | S 4 -- Sn 1 | 0.12 | 2.66494 |
|  | S 4 -- Sn 2 | 0.81 | 2.72711 |
|  | S 2 -- Sn 4 | 0.81 | 2.72711 |
|  | S 3 -- Sn 1 | 0.81 | 2.72711 |
|  | S 1 -- Sn 3 | 0.81 | 2.72711 |

The calculated bond overlap populations and bond lengths between the atoms in the SnS compound are recorded in **Table 16**. From the table, it is observed that the value of the overlap population is small for the given material. The presence of a bonding nature and weak interactions between the atoms are indicated by the small positive value. This results in relatively small elastic moduli and low melting temperature of SnS as found in the previous sections.

### 3.5 Charge Density Distribution

The nature of electronic bonding present in the given SnS system is further investigated with the help of electronic charge density distribution within the (100) and (010) planes. This electronic charge distribution map for the SnS system is shown in **Figure 7**. The color scale present between the two planes (100) and (010) represents the total electron density.

Red on the color scale corresponds to the highest electron density whereas the blue color on the scale corresponds to the lowest electron density. It is observed from the charge density map for SnS that, the charge distribution around Sn and S atoms is almost spherical. It is also observed that there is a high electron density around the S atom compared to the Sn atom. This result agrees with the Mulliken and Hirshfeld charge analysis. Again, in SnS the electron density on the S atom is found much higher than on the Sn atom as the electronegativity of S is greater than that of the Sn atom. This is also a clear sign that an ionic bonding is present between the S and the Sn atoms. From **Figure 7** it can be seen that SnS also shows slight directional bonding between Sn and S where clear signs of charge accumulation between them are seen. This indicates that partial covalent bonding is also present in SnS.

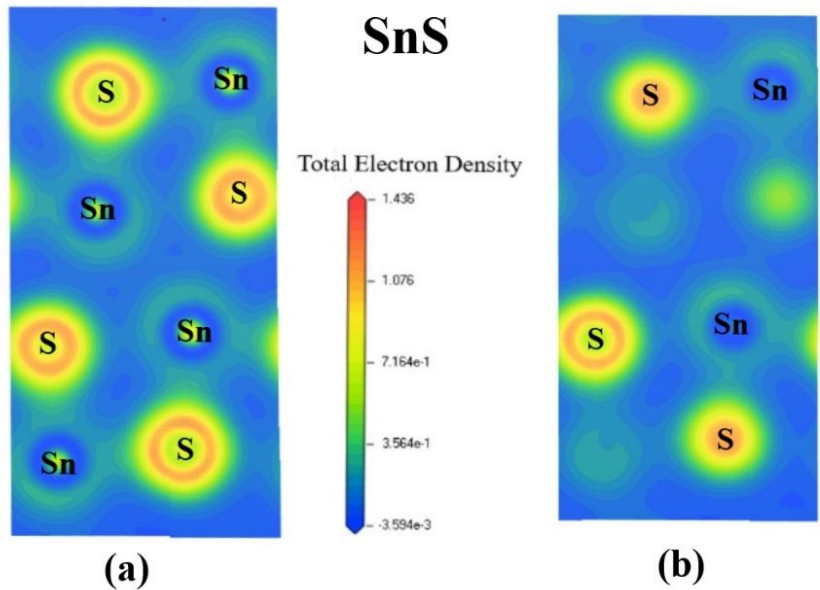

**Figure 7.** Charge density distribution map of SnS in the (a) (100) and (b) (010) planes.

## 3.6 Optical Properties

In recent years, with the increasing advancement and demand for photovoltaic and optoelectronic devices, the study of different energy-dependent optical properties of materials has become a crucial part of fundamental and applied research in materials science. The materials suitable for applications in the different optoelectronic devices are predicted through such studies. The different energy-dependent optical parameters usually studied are namely, dielectric function

ε(ω), refractive index n(ω), optical conductivity σ(ω), reflectivity R(ω), absorption coefficient α(ω), and energy loss function L(ω) (where ω = 2πf is the angular frequency) and the response of these parameters to the incident photon energy were calculated and explored in this section. The complex dielectric function could be obtained from,

$$\varepsilon(\omega) = \varepsilon_1(\omega) + i\varepsilon_2(\omega).$$

The imaginary part and the real part are connected through the Kramers-Kronig relationship. All the other optical constants of interest are determined from the following relationships [83]:

$$n(\omega) = \sqrt{\frac{|\varepsilon(\omega)| + \varepsilon_1(\omega)}{2}}$$

$$k(\omega) = \sqrt{\frac{|\varepsilon(\omega)| - \varepsilon_1(\omega)}{2}}$$

$$R(\omega) = \frac{(n-1)^2 + k^2}{(n+1)^2 + k^2}$$

$$\alpha(\omega) = 2k\omega/c$$

$$L(\omega) = Im\left(\frac{-1}{\varepsilon(\omega)}\right) = \frac{\varepsilon_2(\omega)}{\varepsilon_1^2(\omega) + \varepsilon_2^2(\omega)}$$

$$\sigma(\omega) = \sigma_1(\omega) + i\sigma_2(\omega) = -i\frac{\omega}{4\pi}[\varepsilon(\omega) - 1]$$

Whether a material is a metal, a semiconductor or a non-metal could be easily predicted from the study of the absorption coefficient. The absorption coefficient, *α(ω)* of SnS at 0 GPa pressure for the electric field polarization directions [100], [010], [001] is shown in **Figure 8 (a)**. For the compound SnS, the absorption coefficient started at about ~ 0.50 eV. In this case, the onset energies were in good agreement with the band gap value obtained from the band structure calculations which further enhances the claim that SnS is a small band gap semiconducting material. For SnS, the *α(ω)* was quite high in the regions from about 3.09 eV to 12.94 eV peaking at ~ 9.09 eV.

The loss function of the given compound is shown in **Figure 8 (b)**. The highest peak of the energy-loss spectra is defined as the bulk plasma frequency, $\omega_P$. The loss peak is found at ~

19.19 eV for SnS. At the energy specified by the loss peak, plasma oscillations due to collective motions of the charge carriers are induced. It is also observed that the absorption coefficient and reflectivity spectra have a sharp fall at an energy that coincides with the plasma energy. This implies that the titled compound are expected to behave transparently for photons with energies greater than the plasma energy and the optical properties would show behaviors appropriate for insulating systems in this spectral region.

The reflectivity spectra at 0 GPa pressure are shown in **Figure 8 (c)**. The reflectivity for SnS starts at a value of ~ 0.371 along [100] direction and at ~ 0.481 along both [010] and [001] polarization directions. Therefore, there is optical anisotropy in reflectivity. The reflectivity spectra increased up to ~ 4.7 eV for SnS then fluctuates up to ~ 19 eV. Then it starts to fall rapidly. In the given compound $R(\omega)$ remained over 50% in the energy range from ~2 eV to ~ 19 eV for SnS for the [010] polarization direction. Here, $R(\omega)$ displays almost non-selective behavior. Therefore, SnS can be employed as an efficient reflector of solar radiation.

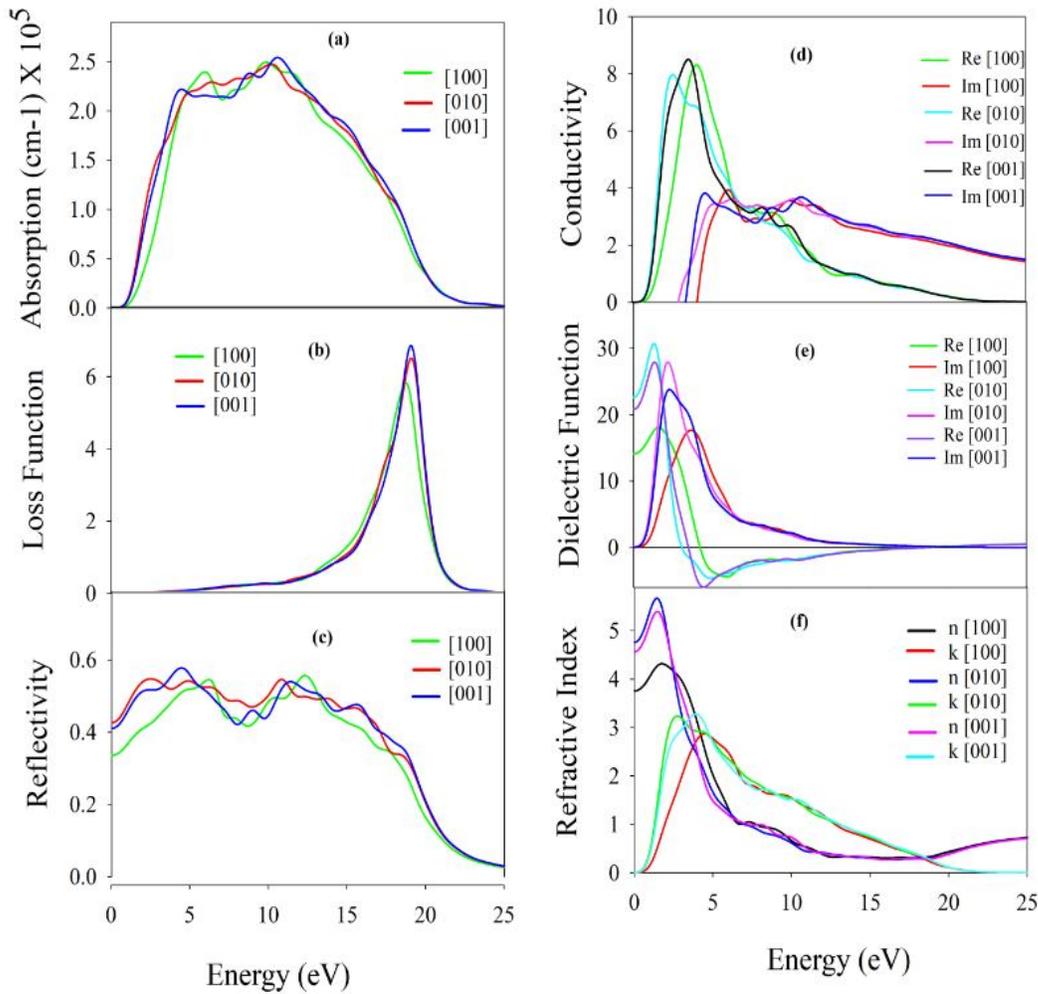

**Figure 8.** The energy dependent (a) absorption coefficient (b) loss function (c) reflectivity (d) optical conductivity (e) dielectric function and (f) refractive index of SnS with electric field polarization vectors along [100], [010], and [001] directions at 0 GPa pressure.

The optical conductivity spectra for SnS are shown in **Figure 8 (d)**. From the figure, it can be seen that the conductivity of SnS starts at about ~ 0.60 eV. This is another proof that the titled material is semiconducting in nature. As the material absorbs the photon energy, their conductivity was found to increase primarily due to the increase in the electron and hole concentrations. The real part of the optical conductivity exhibits the highest peaks at ~ 4.09 eV for [100], at ~2.35 eV for [010] and at ~ 2.50 eV for [001] polarization directions. The imaginary part exhibits peaks at ~5.90 eV for [100], at ~ 5.19 eV for [010], at ~ 4.64 eV for [010] polarization directions.

The real and imaginary parts of the dielectric function are illustrated in **Figure 8 (e)**. From the figure, the real part of the dielectric function is found to increase with energy up to ~ 1.42 eV for [100], up to ~1.03 eV for [010], and up to ~ 1.10 eV for [001] polarization directions of the electric field vector for SnS. Then the real part is found to decrease and becomes negative and again increases finally assuming a low value at ~ 19.5 eV.

**Figure 8 (f)** shows the real and imaginary parts of the refractive index. Because of being a complex quantity, the refractive index has two parts: the real part and the imaginary part. Phase velocity of the electromagnetic wave in SnS is determined by the real part of the refractive index whereas the imaginary part of the refractive index represents the extinction coefficient which gives the measure of attenuation of energy. From the figures, it can be seen that the real part of the refractive index is high in the infrared region and then gradually decreases as it progressed across the visible and ultraviolet regions. The extinction coefficient vanishes at ~ 21.94 eV for SnS. There is optical anisotropy in the real part at low photon energies. High values of refractive index at low energies (containing the visible region) and small semiconducting band gap suggests that SnS is a suitable system for optical display device applications.

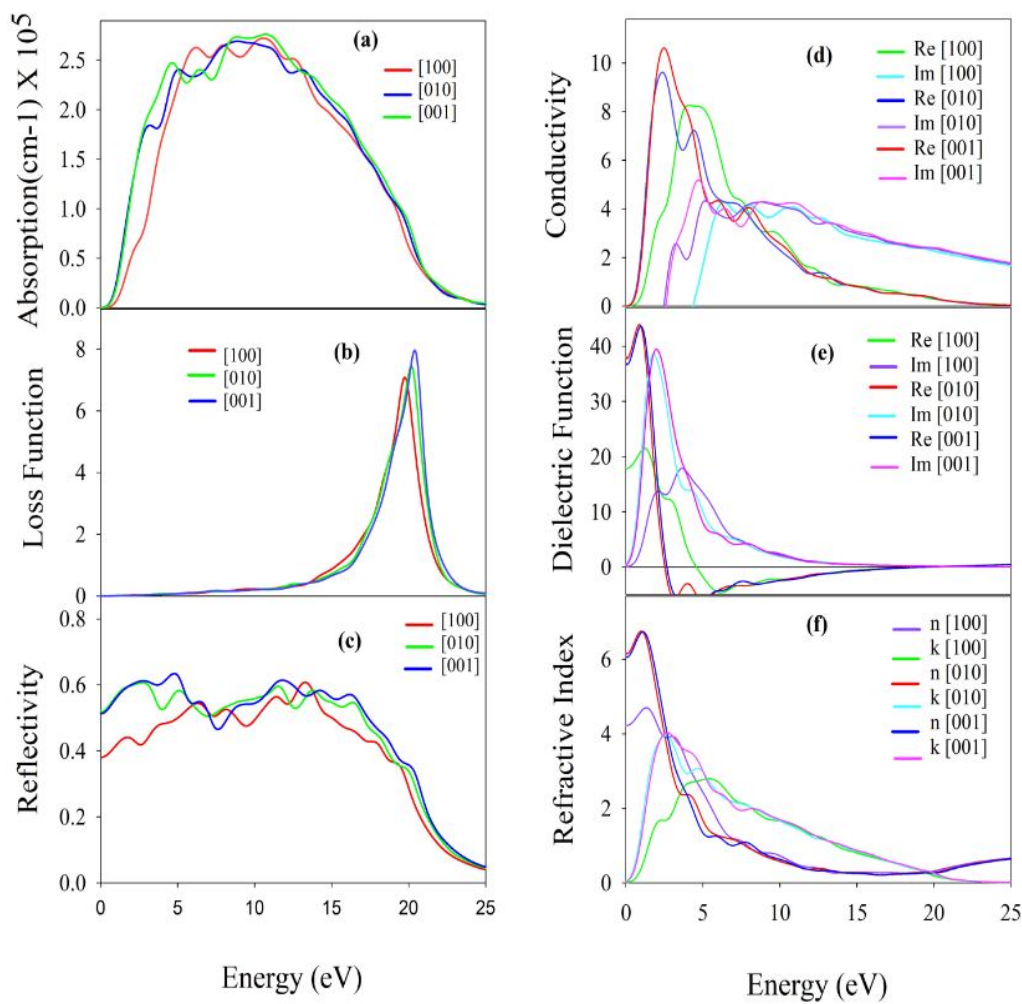

**Figure 9.** The energy dependent (a) absorption coefficient (b) loss function (c) reflectivity (d) optical conductivity (e) dielectric function and (f) refractive index of SnS with electric field polarization vectors along [100], [010], and [001] directions at 5 GPa pressure.

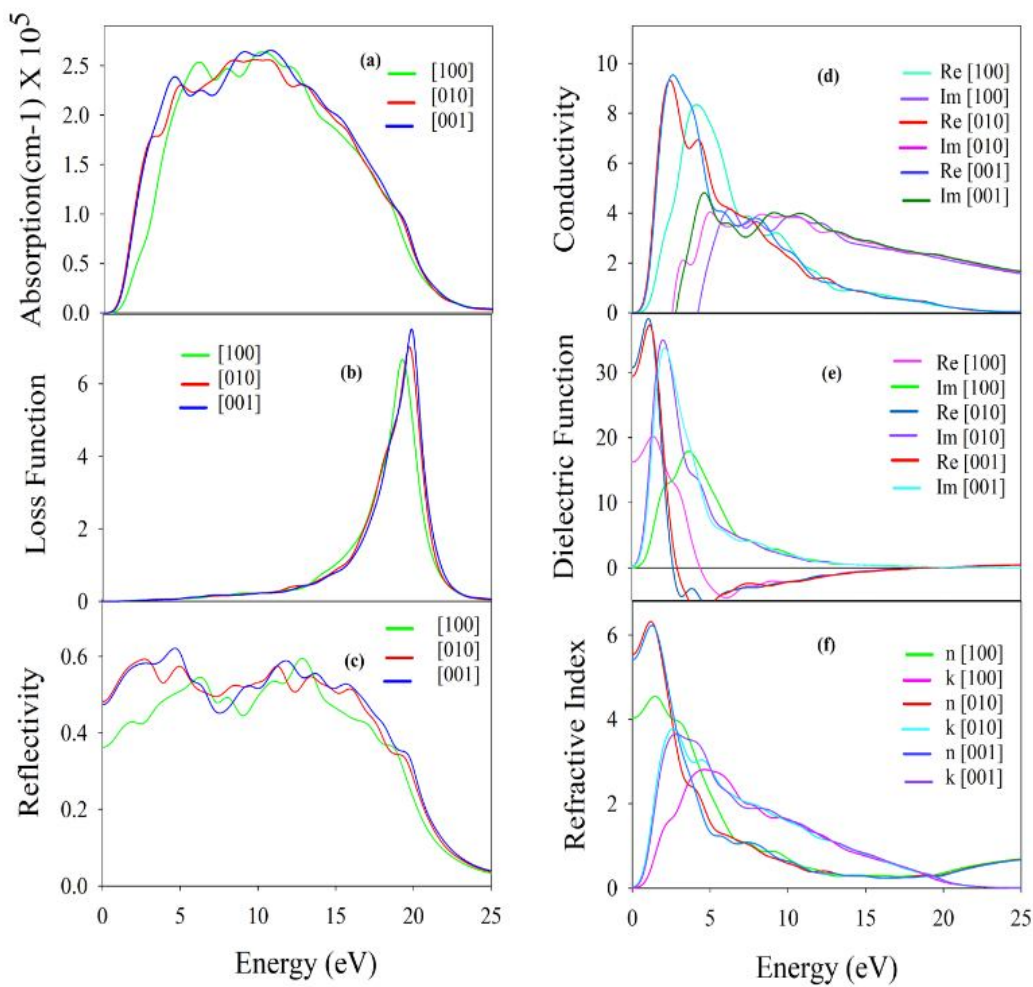

**Figure 10.** The energy dependent (a) absorption coefficient (b) loss function (c) reflectivity (d) optical conductivity (e) dielectric function and (f) refractive index of SnS with electric field polarization vectors along [100], [010], and [001] directions at 10 GPa pressure.

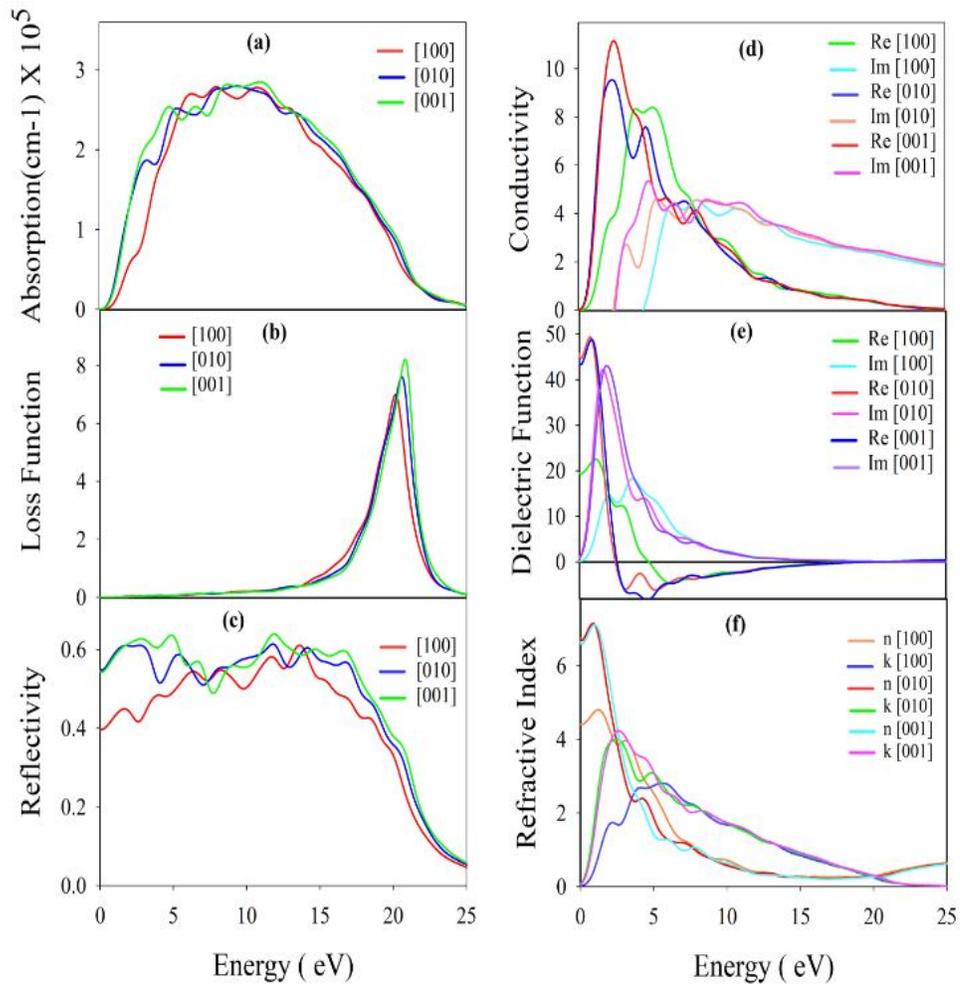

**Figure 11.** The energy dependent (a) absorption coefficient (b) loss function (c) reflectivity (d) optical conductivity (e) dielectric function and (f) refractive index of SnS with electric field polarization vectors along [100], [010], and [001] directions at 15 GPa pressure.

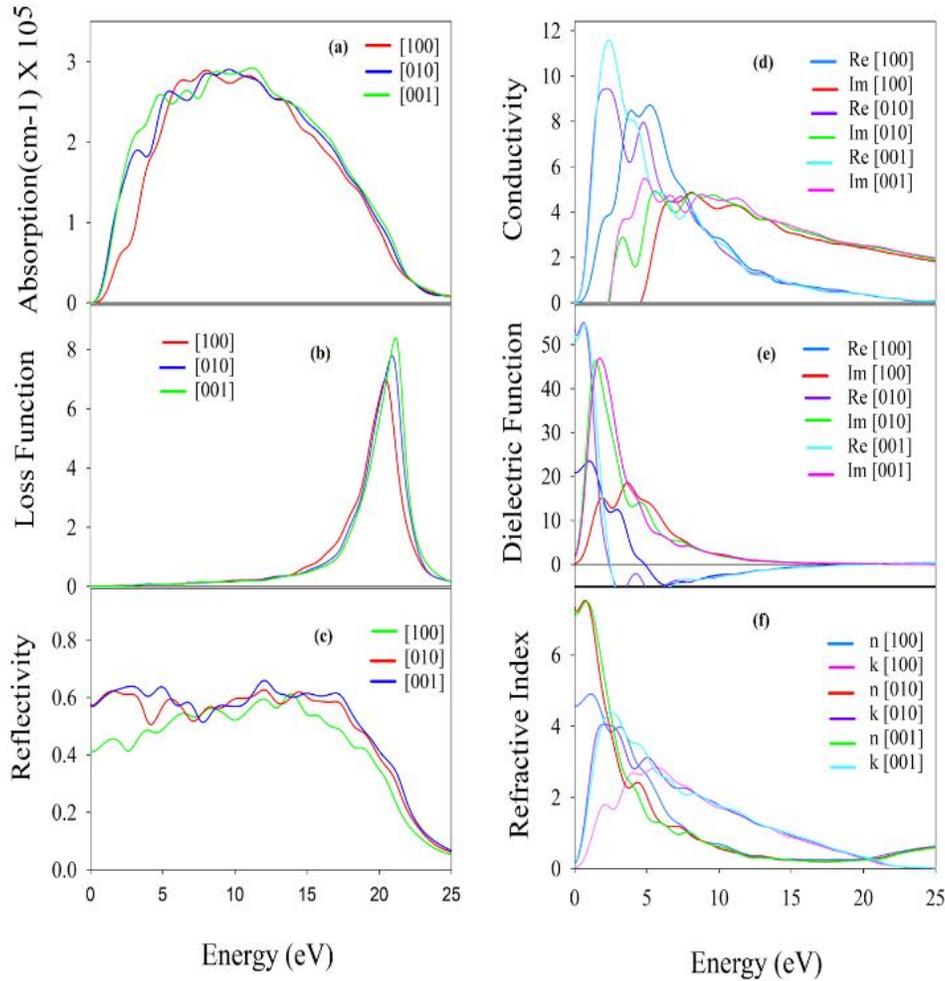

**Figure 12.** The energy dependent (a) absorption coefficient (b) loss function (c) reflectivity (d) optical conductivity (e) dielectric function and (f) refractive index of SnS with electric field polarization vectors along [100], [010], and [001] directions at 20 GPa pressure.

Since this investigation involved the study of the effect of pressure on the titled compound, the optical properties are illustrated in the **Figures 9**, **Figures 10**, **Figures 11**, and **Figures 12** for 5 GPa, 10 GPa, 15 GPa, and 20 GPa pressures, respectively.

From the observation of the figures, it is seen there was no drastic change in the optical properties of SnS due to the increase in pressure, other than metallic characteristics. Namely, the absorption coefficient and the optical conductivity start increasing from zero photon energy. The low energy reflectivity also increases with increasing pressure induced increase in the metallicity. This makes SnS a more efficient reflector of solar radiation when external pressure is applied. The low energy real part of the refractive index also increases with increasing pressure. The loss peak shifts slightly to higher energy with increasing pressure. For all the pressures considered in this work, SnS exhibits high absorption capability of the ultraviolet radiation in the energy range 5 eV to 15 eV, irrespective of the electric field polarization of the incident radiation.

## 4. Conclusions

DFT based study of the pressure dependent physical properties of binary SnS compound has been carried out in this paper. Most the results presented herein are novel. The results revisited are found to be in good agreement with previous reports. The elastic constants and moduli of SnS show that it is mechanically stable at all the pressures considered. The compound is brittle under ambient conditions but undergoes brittle-ductile transformation under pressure. This is probably related to the increasing metallicity of SnS with increasing pressure. In the ground state SnS is a direct band gap semiconductor with an energy gap ~ 0.60 eV. The compound becomes metallic under pressure. The pressure induced increase in the metallicity, decrease in the Coulomb pseudopotential, and the increase in the Debye temperature are expected to facilitate superconductivity in SnS under high pressure. Various anisotropy indices reveal that the compound is elastically/mechanically anisotropic. Both Debye temperature and lattice thermal conductivity are low in the ambient conditions but both these parameters increase significantly with pressure. The Melting temperature of the SnS is quite low as manifested by moderate hardness of the compound in the ground state. From MPA and HPA, mixed bonding characteristics are found with ionic and covalent contributions in SnS. The optical parameters are investigated theoretically in detail. The energy dependent optical parameters show very good agreement with the electronic band structure calculations. SnS is an efficient absorber of ultraviolet light. The reflectivity of SnS increases with the increase in pressure. The reflectivity spectrum is fairly nonselective over a wide spectral range including the visible region. The low energy refractive index is high at all pressures considered in this study. All these optical features can be useful for potential optoelectronic device applications.

**Acknowledgments**
S.H.N. and R.S.I. acknowledge the research grant (1151/5/52/RU/Science-07/19-20) from the Faculty of Science, University of Rajshahi, Bangladesh, which partly supported this work.

S.H.N. dedicates this paper to the loving memory of his father, Professor A.K.M. Mohiuddin, who passed away this year.

**Declaration of interest**

The authors declare that they have no known competing financial interests or personal relationships that could have appeared to influence the work reported in this paper.

**Data availability**

The data sets generated and/or analyzed in this study are available from the corresponding author on reasonable request.

**CRediT author statement**

**Ayesha Tasnim:** Methodology, Software, Writing- Original draft. **Md. Mahamudujjaman**: Methodology, Software. **Md. Asif Afzal**: Methodology, Software. **R.S. Islam:** Supervision, Writing- Reviewing and Editing. **S.H. Naqib:** Conceptualization, Supervision, Formal analysis, Writing- Reviewing and Editing.